# WHAT IS NEEDED FOR BISCO TO WORK IN A DIPOLE INSERT FOR 20 TESLA HYBRID ACCELERATOR MAGNETS *

E. Barzi†, Fermi National Accelerator Laboratory, Batavia, USA
†, also at Ohio State University, Columbus, USA

*Abstract*

A major focus of the global HEP community is on high-field superconducting magnets made of High Temperature Superconductors (HTS) for future Energy Frontier Programs. Within the U.S. Magnet Development Program (US-MDP), a key task is that of developing HTS inserts producing fields larger than 5 T within 15 T outserts made of $Nb_3Sn$ to generate 20 T+ for future accelerators. $Bi_2Sr_2CaCu_2O_{8-x}$ (BiSCO) is the only high-$T_c$ superconductor available as an isotropic round multifilamentary wire, which is ideal for producing the flat cables (i.e., Rutherford-type cables) that are used in accelerator magnets. Significant progress in the development and industrialization of BiSCO wires has been made over the last decade. However, several challenges remain for this HTS to be used successfully in hybrid magnets. The following is required to improve performance, lower costs and simplify the processing of BiSCO accelerator magnets: (1) The development and design, in collaboration with industry, of BiSCO wires that are adequate for Rutherford cabling; (2) The development of insulation processes and materials that prevent leaks when heat treated in highly corrosive oxygen; (3) Control of stresses and strains; and (4) Integration of high-pressure heat treatment with a new approach that will lower costs and simplify processing.

## INTRODUCTION

The 2014 Particle Physics Project Prioritization Panel (P5) strategic plan for U.S. High Energy Physics, echoed by the HEPAP subpanel review of the General Accelerator R&D (GARD) program [1], endorsed a continued world leadership role in superconducting magnet technology for future high energy colliders. The U.S. Magnet Development Program (US-MDP) [2-4] was created in 2016, with a major goal to develop HTS inserts that produce magnetic fields larger than 5 T within 15 T $Nb_3Sn$ outserts to generate 20 T or higher fields.

Developing accelerator magnets up to or exceeding 20 T requires both high temperature superconductors (HTS) and Low Temperature Superconductors (LTS). The continued progress toward higher magnetic fields holds significant potential for general advances in science and technology [5]. For the past twenty years there have been steady increases in the field strength of superconducting magnets thanks to constant development in $Nb_3Sn$ conductor and magnet technology. In 2009, internal tin $Nb_3Sn$ conductors have enabled a commercial 23.5 T magnet for 1 GHz NMR spectroscopy [6], with an actively shielded version produced in 2016 [7]. In 2020, Bruker Corporation announced the successful installation and customer acceptance of the world's first Avance™ NEO 1.2 GHz NMR system at the University of Florence, Italy [8]. Bruker's NMR magnets with 54 mm bore utilize a novel hybrid technology with HTS in the inner sections, and LTS in the outer sections. In accelerator technology, following the 1997 record magnetic field at 4.5 K of 12.8 T (13.5 T at 1.9 K) in a four-layer cos-theta D20 LBNL dipole, in 2019 the FNAL team achieved a world record field at 4.5 K of 14.1 T in an accelerator dipole [9], pushing it up to 14.6 T at 1.9 K in 2020 [10].

Practical and commercially available HTS include round $Bi_2Sr_2CaCu_2O_{8-x}$ (BiSCO) composite wires and ribbons of rare-earth barium copper oxide (ReBCO). Round BiSCO composite wires (Fig. 1, left) have the key advantage that they can be directly applied to the flat Rutherford cables (Fig. 1, right) cost-effectively used in high-field accelerator magnets [11,12]. Significant advances were made in the development and industrialization of BiSCO. Km-length quantities of BiSCO composite round wires are commercially produced. The critical current density $J_c$(20 T, 4.2 K) and the engineering current density $J_e$(20 T, 4.2 K) of the round wire increased from 1500 A/mm$^2$ and 400 A/mm$^2$ or less respectively in the 2000s, to $J_e$ values greater than 800 A/mm$^2$ in the best produced wire samples today. The two main factors that led to this improvement are: (1) The removal of 30% porosity in as-drawn BiSCO wires by an overpressure processing heat treatment at 50 bar; and (2) The introduction of a new chemical powder technology by Engi-Mat, which produces highly homogenous BiSCO precursor powders with very good composition control. Nevertheless, BiSCO is not yet a magnet-ready superconductor. Several challenges remain to be addressed in the areas of both materials and magnet technology.

Composite BiSCO is a soft and very delicate high-$T_c$ copper oxide, which needs rigorous empirical laws for Rutherford cabling to minimize internal damage. Once the cable is formed and used to wind a coil, a multistage heat treatment in a highly corrosive oxygen ($O_2$) atmosphere at maximum temperatures close to 900°C is required. Temperature homogeneity has to meet stringent gradient specifications. The BiSCO sometimes leaks through the encasing Ag alloy metal. Leaks degrade the $J_e$ locally [13]. This problem is related to both the strand ability to withstand deformation and the chemical compatibility of the insulation material with the Ag alloy during heat treatment in $O_2$.

BiSCO wires and cables are sensitive to strain [14-17]. Although BiSCO is universally made with a technique called Powder-in-Tube (i.e., the wire subcomponents are Ag tubes filled with BiSCO powders), the strain behaviour appears to depend on the different Ag alloy used by each manufacturer for the outer sheath of the wire.

___________________
* Work supported by Fermi Research Alliance, LLC, under contract No. DE-AC02-07CH11359 with the U.S. DOE and the U.S. Magnet Development Program (US-MDP).
† barzi@fnal.gov.

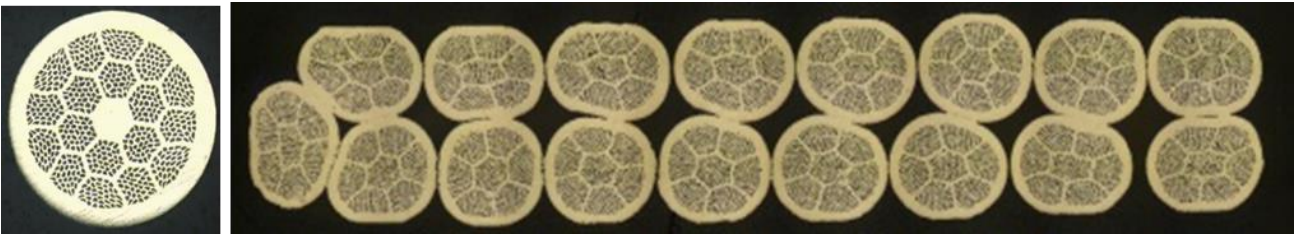

Figure 1: Cross section of BiSCO round composite wire (left) and of BiSCO Rutherford cable (right).

While very accurate measurements on tensile strain have been made, the main stress component in accelerator magnets, i.e. the coil azimuthal pressure, is transverse to the cable. More data on BiSCO performance sensitivity to transverse cable loading are therefore vital. Stress management concepts need to be applied to the design of insert coils when aiming at large magnetic fields of hybrid magnet systems. Coil test, quench detection, and protection also need to be addressed.

A BiSCO dipole insert for an accelerator magnet is made of insulated Rutherford cable wound inside a mechanical support structure. $O_2$ access to the BiSCO superconducting volume might be limited and require appropriate solutions. This is the case for both the Canted Cosine Theta (CCT) coil concept [13] by the LBNL group, and FNAL coil inserts based on a cos-theta coil design with stress management elements [18]. The minimum acceptable coil aperture is given by the mechanical properties of the BiSCO Rutherford cable that has to be bent in the plane of its flat face at the coil mid-plane, and in a plane perpendicular to it at the poles. Refined magnetic and mechanical analyses are performed with simulation software such as ROXIE and ANSYS. After winding, the actual coil is heat treated at 50 bar in unique facilities.

## OBJECTIVES

New and robust BiSCO wires for the cable geometry required in accelerator magnets have to still be developed by industry. New ancillary materials and processes have to be explored, characterized and tested. Solutions to the challenges that BiSCO still presents as a magnet conductor [3] have to be pursued. Endeavors include the following.

### Eliminate Bisco Leaks

Heat treating BiSCO at high gas pressure does not completely eliminate leaks from the Rutherford cable. Any cable defects, such as leaks of melted BiSCO, are unacceptable in accelerator magnets. Superconducting leaks are a major reason for the loss of critical current in coils. Superconducting round wires are obtained by further processing larger composite cylinders called billets.

New and improved billets have to be designed in collaboration with Bruker-OST LLC (BOST) and any other capable industry to produce wires that are adequate for Rutherford cabling. Options include using a thicker AgMg sheath and/or a thicker Ag web between the superconducting filaments. A thicker Ag web design helps the wire to survive large and uneven deformations during the cabling process. A thicker Ag web is also expected to produce a larger $J_c$ in the wire, thanks to the more homogeneous microstructure. This increased $J_c$ performance will compensate for critical current loss due to the reduction of the fill factor in the new designs. In addition, because a more robust wire will not leak, any decrease in the $I_c$ of the round wire will be compensated with much less degradation of the $I_c$ in the cable, and therefore in the magnet itself. A thicker Ag web is expected to also produce wires whose $J_c$ is less sensitive to the maximum melting temperature. This is very important, as also shown in [19-21].

To test the applicability of composite wires to cable fabrication, BiSCO wires can be first flat-rolled to increasing deformations to systematically study such dependence of their properties. By leveraging existing FEM models realized for $Nb_3Sn$, simulations of this mechanical process can be reproduced for BiSCO wires, and a sensitivity analysis to the main parameters can help in optimizing the wire design.

### Test New Bisco Billets in Rutherford Cable

Once an adequate BiSCO wire has been developed for cabling, wires extracted from Rutherford cables of appropriate geometry can be tested by measuring their superconducting properties at the FNAL Superconducting R&D lab.

Rutherford cables can be tested at self-field at FNAL with a superconducting transformer up to ~ 20kA [22]. Cable $I_c$ measurements can also be performed at FNAL in an external magnetic field with a Rutherford cable test facility [23,24] with bifilar sample and superconducting transformer that operates in a 14 T/16 T Teslatron system by Oxford Instruments. The power supply for the NbTi primary coil is 875 A at 5 V. The secondary winding is equipped with a heater to quench the current in the secondary before each primary new excitation step, and with a Rogowski coil to measure the secondary current. The secondary current is corrected from the linear contribution due to the primary stray field, and in $Nb_3Sn$ cable tests it has reached values > 25 kA. To test BiSCO cable samples, the appropriate Inconel-600 parts required for the bifilar sample holder need to be procured.

### Develop Chemically Compatible Insulation

To realize an effective accelerator magnet, an insulation material chemically compatible with BiSCO and its high temperature processing in $O_2$ is necessary and still has to be found. Existing methods using mullite sleeve and $TiO_2$- polymer slurry [13] have not shown to completely eliminate leaks, due to the abrasion of the $TiO_2$ coating on the bare cable. Further investment in research on compatible insulation without Silica and associated processes is needed to solve this problem permanently.

Experimental studies need to be performed on insulated BiSCO cable, wound at different bending radii, and then heat treated in $O_2$ within a structure of Inconel-600 and/or of any other structural material. Such experiments allow (1) systematically measuring any produced leaks; (2) obtaining the minimum bending radius that is acceptable for a given BiSCO cable geometry; and (3) testing chemical compatibility of the whole BiSCO coil package materials. Included wire samples, whose $I_c$ is tested after reaction, are used as witnesses of the reaction quality.

In addition, impregnation materials other than the CTD-101 and FSU Mix 61 currently used within US-MDP, with better mechanical and radiation resistance properties, have to be investigated.

### Strengthen Bisco Wires, Cables and Coils

BiSCO wires and cables are strain sensitive. To use BiSCO coils as inserts in very high field magnets, the stresses and strains must be controlled and limited as BiSCO is more sensitive to stress than $Nb_3Sn$. This can be done by acting on the following fronts:

a. Invest in research and development of methods to mechanically reinforce the wire (as in the first Objective above), but also the Rutherford cable itself. One example is to wrap the cable with appropriate metal ribbons.

b. Use coil stress management elements to reduce stress in the insert coils while also being chemically compatible with the BiSCO processing. Leverage ongoing dipole insert design studies and optimization performed by magnet teams using Finite Element analysis.

c. Monitor progress through accurate $I_c$ measurements of BiSCO cable samples under transverse pressure. This can be done up to about 200 MPa on impregnated cable samples with the FNAL Transverse Pressure Insert (TPI) fixture. This device produces the effect of uni-axial and not multi-axial strain, since the experimental setup allows for the sample to expand laterally.

### Integrate High-Pressure Heat Treatment and Split-Melt Approaches

To lower costs and simplify the processing of BiSCO inserts for hybrid accelerator magnets, the Split Melt Process (SMP) [25, 26] needs reconsidering. SMP is the heat treatment of the BiSCO when split into two separate heat treatments and the coil is wound between them. The very first studies produced promising results on round wires, with straight, undeformed wire samples seeing a 40% increase in $J_c$, and samples bent at 15 mm radius after the first stage of the heat treatment, seeing a 30% increase in $J_c$. A hybrid approach where one or both cycles of the Split Melt Process are carried out at 50 bar has never been explored experimentally. In the assumption that the ceramic leakage and $J_c$ sensitivity to peak times and temperature gradients are more critical at the BiSCO melting temperature, either loosely wound wire or even just Rutherford-type cable wound on a spool would be heat treated at 50 bar for the first stage of the thermal cycle to reduce as much as possible the volume requiring maximum sensitivity. Another important benefit of heat treating a spool of loosely wound cable as opposed to a tightly wound coil might be that of reducing the ceramic leaks that occur at the liquid stage, and of a more homogenous $O_2$ diffusion.

A concern on the effectiveness of the proposed hybrid method was based on the concept that the bi-axial grain alignment that occurs during the cooling from the BiSCO melting temperature and that is required for a good $J_c$ would break during bending [27,28]. Preliminary $I_c$ results obtained at FNAL on BiSCO wire samples wound at 6.6 mm radius after the first stage of the heat treatment [3] are proving that phases that form after the cooling down to room temperature from the melting temperature are not strain sensitive.

### Apply Best Materials and Processes to Bisco Prototypes

Selection of the most promising BiSCO wires and processes has to be made on the basis of experimental characterization. The best obtained coil technology procedures will be used for BiSCO sub-scale coils from LBNL and FNAL within US-MDP. Fast turnaround tests that make use of only one coil, i.e., half a magnet, can be tested in the existing FNAL 11 T dipole mirror structure with 60 mm aperture, and soon in a planned 17 T dipole mirror structure. In a mirror test configuration, a half-coil is tested with the other half being replaced with bulk iron, which reflects the magnetic field produced by the half-coil. In this configuration the total field and Lorentz forces are only slightly smaller and therefore it is a faster and less expensive, albeit sufficiently representative test, in lieu of testing an entire coil.

## CONCLUSION

The HEP global community has been ushering in a new era of high-tech accelerator development also through the strong endorsement of the European Strategy for Particle Physics of "high-field superconducting magnets, including high-temperature superconductors" [29]. For instance, CERN's Magnet group is investing $100M+ over 5 years on this topic for Materials&Services alone. The proposed research would help US-MDP sustain the U.S. world leadership position in HEP accelerator magnets by making BiSCO magnet-ready. Results can be also applied to very high field wigglers to improve the output and capabilities of light sources for Basic Energy Sciences, to superconducting magnets for plasma confinement in Fusion Energy Sciences, and to commercial high field magnet systems for high frequency NMR spectroscopy. According to the Bruker Corporation, the development of a BiSCO wire suitable for NMR applications is expected to reach a demand greater than 1 ton/year by 2030 [30].